# Privacy Preserving Anomaly Detection on Homomorphic Encrypted Data from IoT Sensors

Anca Hangan, Dragos Lazea, Tudor Cioara*

Computer Science Department, Technical University of Cluj-Napoca, Memorandumului 28,

400114 Cluj-Napoca, Romania

{firstname.lastname}@cs.utcluj.ro

**Abstract.** The Internet of Things (IoT) devices have become indispensable components of our lives, and the advancement of AI technologies will make them even more pervasive, increasing the vulnerability to malfunctions or cyberattacks and raising privacy concerns. Encryption can mitigate these challenges; however, most existing anomaly detection techniques decrypt the data to perform the analysis, potentially undermining the encryption protection provided during transit or storage. Homomorphic encryption schemes are promising solutions as they enable the processing and execution of operations on IoT data while still encrypted, however, these schemes offer only limited operations, which poses challenges to their practical usage. In this paper, we propose a novel privacy-preserving anomaly detection solution designed for homomorphically encrypted data generated by IoT devices that efficiently detects abnormal values without performing decryption. We have adapted the Histogram-based anomaly detection technique for the Fast Fully Homomorphic Encryption over the Torus scheme to address limitations related to the input size and the depth of computation by implementing vectorized support operations. These operations include addition, value placement in buckets, labeling abnormal buckets based on a threshold frequency, labeling abnormal values based on their range, and bucket labels. Evaluation results show that the solution effectively detects anomalies without requiring IoT data decryption and achieves consistent results comparable to the mechanism operating on plain data. Also, it shows robustness and resilience against various challenges commonly encountered in IoT environments, such as noisy sensor data, adversarial attacks, communication failures, and device malfunctions. Moreover, the time and computational overheads determined for several solution configurations, despite being large, are reasonable compared to those reported in existing literature.

## 1. Introduction

In today's digital era, the Internet of Things (IoT) devices have become indispensable components of our lives. They have become pervasive and integrated with every activity, from smart controllers, meters, or wearable fitness trackers to interconnected homes and smart cities, shaping our interaction with our surroundings, and impacting modern society's development [24], [25]. Each IoT sensor is usually connected to a platform that integrates monitored data from the devices and applies analytics to identify patterns, facilitate decision-making and suggest recommendations [26]. Moreover, the maturation of AI technologies changes the landscape of IoT field transforming the devices from simple transmitters of monitored data into intelligent machines capable of understanding the data, take decisions and coordinate their execution with the user needs [27]. As IoT devices generate large amounts of data at a high frequency to support the control processes, anomaly detection plays an important role in IoT and AI, facilitating proactive interventions and improving reliability while enabling various applications across different domains [28]. Anomaly detection focuses on identifying unexpected data points or data patterns generated by IoT devices being is essential to prevent disruptions and maintain the integrity of critical large-scale infrastructures [29]. Heterogeneous and distributed IoT devices continuously monitor and collect data from various aspects of our lives, will generate vast amounts of data useful to optimize services such as healthcare or transportation of energy delivery [27][30].

However, the influx of IoT data at high velocity from heterogenous devices also increases the vulnerability to cyberattacks and devices malfunctions. Ensuring the security and privacy of this data is paramount, as any breach can have serious consequences, including the compromise of critical infrastructure operation and leaks of sensitive citizen information [31]. Traditional anomaly detection methods often involve analyzing raw data directly, which may pose privacy risks, especially when dealing with sensitive or personal data. Thus, privacy preserving anomaly detection techniques are proposed to enable the analysis of IoT data streams without revealing the sensitive data [32], [33]. They leverage obfuscating personally identifiable information, adding noise or perturbation to the data to hide personal data, or encryption using cryptographic methods. The encryption-based privacy-preserving anomaly detection techniques offer robust protection of individuals' privacy while enabling the analytics to derive valuable insights from data [9][14]. However most existing techniques protect data in transit or in storage, but for anomaly detection procedures the data is decrypted, exposing the raw data to privacy concerns. Therefore, anomaly detection must be performed on a trusted node, that owns the encryption key, but still, raw data is exposed during processing [34], [35].

In this context, homomorphic encryption schemes are promising solutions as they enable the processing and execution of operations on IoT data while still encrypted. As a result, they may ensure that the raw data remains encrypted throughout the entire anomaly detection process as well as the safe sharing of encrypted data among peer organizations to increase the accuracy of the process [34]. However, there are limitations on the type of operations (only addition, multiplication are widely supported by HE schemes) and the range and type of the operands (usually integer values) [3, 4, 6, 10, 12]. Difficulties related to secure computations such as comparisons, divisions or exponentiation over the encrypted domain still need to be addressed [32,33]. Overall, encryption/decryption mechanisms and HE data processing generate very large computational overheads that may negatively influence the system's responsiveness making it difficult to address the trade-off between privacy and system operation [35]. Moreover, anomaly detection on homomorphic encrypted data is still an open research problem, and there are only a few attempts toward finding operational solutions. Due to the lack of available operations for encrypted data, the anomaly detection algorithms may not be directly applicable [5], [10]. The selection of the anomaly detection algorithm for IoT data is strictly related to the preponderance support for additive operations support provided by homomorphic encryption schemes [3][11][15]. Even in the case of fully additive anomaly detection schemes, significant challenges need to be addressed, as significant modifications are needed to be compatible with homomorphic encryption [13]. Finally, noise induced by the encryption scheme can affect the anomaly detection accuracy, especially when dealing with small deviations from the IoT device's normal raw data values [11][14].

In this paper, we address the identified challenges in the area by developing a novel solution for anomaly detection in homomorphically encrypted data generated by IoT devices. We leverage the Fast Fully Homomorphic Encryption over the Torus (TFHE) scheme implemented in Concrete [40] to perform operations on encrypted data and use a proven histogram-based technique to identify anomalous sensor readings. The use of the TFHE scheme offers several advantages, such as higher flexibility in implementing various kinds of data analysis procedures, achieving efficiency by translating basic operations into table lookups, and maintaining privacy by ensuring that only the user holds the private key while server-side operations are performed using the public key. We employ a redesigned histogram-based algorithm for homomorphically encrypted data streams from IoT devices to identify which sensor readings are statistically different from previous readings and therefore anomalous. We have implemented the operations of the Equi-Width Histograms anomaly detection technique adapted for the TFHE scheme implemented in Concrete by vectorizing the operations. Our privacy preserving anomaly detection mechanism has consistent results with the corresponding mechanism that operates on plain data, without needing to decrypt the results for decision taking. Moreover, the computational overheads are kept within reasonable limits, making the defined approach a significant contribution to the field.

The novel contributions of the paper are:

- The design of a privacy-preserving anomaly detection technique for homomorphically encrypted data generated by IoT devices that efficiently detects abnormal values based on the frequency of occurrence recorded on previous days, without performing decryption.
- Adaptation of the Histogram-based anomaly detection technique for the TFHE scheme implemented in Concrete by addressing limitations such as the input size and the depth of computation through vectorizing support operations such as addition, value placement in buckets, labeling abnormal buckets based on a threshold frequency, labeling abnormal values based on their range and bucket label.
- Analysis of the technique's performance and computational overhead tradeoffs across different configuration setups and their impact on the overall effectiveness of the privacy-preserving approach within IoT environments.

The rest of the paper is structured as follows. Section 2 presents the related work on privacy-preserving anomaly detection focusing on data encryption and federated data directions. Section 3 provides an overview of the basics of fully homomorphic encryption and the advantages offered by the TFHE encryption scheme. Section 4 presents our solution for anomaly detection on homomorphic encrypted data, and Section 5 presents the evaluation results considering IoT energy metering data. Section 6 discusses our solution's computational and time overheads, and Section 7 concludes the paper and presents the limitations of our work.

## 2. Related work

The state-of-the-art literature follows two research directions to ensure that sensitive data remains private throughout the anomaly detection process. The first direction involves IoT data movement to centralized cloud-based locations and its encryption using cryptographic methods such as homomorphic encryption or differential privacy. The second direction involves federated data processing, which keeps the data local in the proximity of its collection to ensure its privacy.

In the first direction, most of the literature approaches focus on protecting data by combining encryption and various security protocols [4,6]. Their goal is to overcome the impossibility of performing all operations encountered in anomaly detection algorithms, on encrypted data [10,11]. These security protocols are put in place because the data need to be partial or total decrypted to compute comparisons, divisions, exponentiation, and other complex operations [7,14]. However, these protocols also introduce additional overhead to such anomaly detection techniques. Zhang et al. [3] propose an anomaly detection technique based on instance densities, with the assumption that each data provider computes their own densities. They share them in the form of (key, value) pairs, after introducing random disturbance to hide the key and encrypting the value with their private keys. Multiple data sets containing different owner's densities are merged on a centralized server. To detect anomalies, the merged encrypted data is sent back to the owners. The type of encryption used allows homomorphic additions but doesn't support the other operations needed for deciding if an instance is an anomaly or not. Because data owners need help from the server to decrypt the aggregated densities, security protocols between data owners and the server are put in place, resulting in increased network traffic and computational overhead. Alabdulatif et al. [4] define an anomaly detection model based on a private server that will encrypt data received from end users and execute the operations that need decryption, as well as multiple collaborative public servers to perform operations using a homomorphic encryption scheme. The anomaly detection result will be computed by the private server and communicated in a decrypted form to the end user. In this way, the proposed model addresses the need for executing various mathematical computations such as division and comparison, that cannot be executed on an encrypted data domain. The model privacy preserving anomaly detection is extended for decision-making in smart cities [5], where a computational process distribution scheme promoting parallelism is introduced to overcome computational overheads associated with homomorphic encryption. Chen et al. [6] count anomalies in encrypted data using a threshold-based homomorphic encryption scheme. Data providers outsource their data to multiple edge nodes through a secure additive secret-sharing protocol. The edge nodes cooperate to implement a secure windowed Gaussian anomaly detection method with a series of subprotocols. The edge nodes count the anomalous data according to the metrics set by the data requester. Operations needed by the anomaly detection algorithm but unavailable in the homomorphic encryption scheme are implemented by using partial or total decryption.

The blockchain is considered a as a communication and security protocol for anomaly detection exploiting its features such as immutable record keeping, secure data sharing or decentralization [7], [14], [13]. Song et al. [7] propose an anomaly detection service for transactions in a blockchain network. The transactions' feature vectors are sent to a server encrypted and perturbed. The server performs anomaly detection through a k-NN-based method. Distances needed by the anomaly detector are computed on the encrypted features, but the distance vectors need to be decrypted to decide which transaction is abnormal. Shen et al. [14] record on the blockchain the preprocessed and encrypted data. A data analysis node trains SVM models for anomaly detection, performs the checks and updates original data in the blockchain. The comparison operations are implemented through decryption, while addition and multiplication operations, using a homomorphic encryption scheme. Similarly, Mehnaz et al. [13] propose a lightweight and aggregation-optimized encryption scheme that allows for homomorphic addition. All other operations that are needed for anomaly detection are performed by the data owner, in clear, by decrypting the partial results. The experiments show that the proposed privacy-preserving anomaly detection mechanism has practical computation and communication overheads without compromising the results.

Several cryptographic techniques are used in combination to leverage on each other advantages to design robust and privacy-preserving anomaly detection systems [10], [8], [15]. Li et al. [10] define an anomaly detection solution for smart grids in which the operations that are not supported in a homomorphic encryption scheme are performed by a trusted decryptor. Data is encrypted inside smart meters, aggregators at the edge perform additions and multiplications, while a central server solves other operations with the help of a decryptor and lookup tables. Between the server and the decryptor, the secret is preserved by using private information retrieval queries that allow a user to retrieve a record from a database server without letting the server learn which element is selected by the user. Computation latency and communication latency are relatively large. The work is extended in [11] by storing the lookup tables separately to enable arbitrary arithmetic calculations over fully homomorphic encryption and to reduce the execution time of the anomaly detection service while protecting private information. Lai et al. [8] rely on secure multi-party computation to propose a privacy-preserving anomaly detection protocol on incremental data sets, such as network traffic and system logs. The protocol decomposes the anomaly detection algorithm into several phases and recognizes the necessary cryptographic operations in each phase. Calculations are performed using garbler circuits. The experiments show large overheads for communication and processing (execution time and consumed memory) during model initialization and updates. Finally, Alexandru et al. [15] propose a privacy-preserving anomaly detection technique for linear control systems implemented using garbler circuits, homomorphic encryption, and a combination of the two. State estimations for control operations need matrix multiplication. Anomaly detection is implemented with a cumulative sum algorithm that needs comparisons that are very challenging to implement on homomorphically encrypted data. Due to the computational overhead associated with homomorphic encryption and communication overheads of secured multi-party computation, a hybrid solution is proposed. The state estimations are computed on homomorphic encrypted data, while the cumulative sum algorithm is implemented using garbled circuits. The hybrid approach significantly reduces overheads and thus can be considered a feasible solution.

The federated data solutions for anomaly detection keep data and perform processing locally, thus eliminating the possibility of untrusted parties handling unprotected data. Since they only need the exchange of aggregated, communication costs and latency can be lower. However, in some cases, there is a significant communication overhead involved if the anomaly detection model is large and the synchronization among parties needs to occur frequently [8, 15]. Moreover, because multiple parties collaborate in the anomaly detection process, there is a need for interoperability rules and additional security measures for transferring and storing aggregated data [12, 17]. Due to the drawbacks of privacy-preserving data processing techniques, hybrid models integrating cryptographic techniques are more likely to lead to practical solutions and feasible deployments [16, 18, 20]. There are a few methods that combine federated learning for anomaly detection and local differential privacy [17, 18] or encryption [12] to obfuscate the local models' parameters that are shared with a central model aggregator in the IoT system. Differential privacy overhead is introduced by the data alteration procedure for data extraction from secured storage to transferred or processed. However, it is lower compared to encryption [9, 18]. The downside of this technique is that data can be compromised by multiple queries on the same data set or if the party doing the processing has additional information. Truex et al. [18] show that only a certain number of queries are permitted before the secret is compromised. Moreover, the perturbation introduced by the differential privacy mechanism can reduce the accuracy of the federated learning model. When using encryption, on the other hand, all data owners need to share the same key, and parameter aggregation needs to be performed, generating overheads due to key sharing and cryptographic operations. Wen et al. [9] detect energy theft in smart grids using federated learning that is performed on detection stations closer to the Edge. Data received from multiple consumers is kept private on detection stations through differential privacy. Training parameters are encrypted, sent to the cloud aggregated using a homomorphic encryption scheme, and distributed back to the detection stations. Itokazu et al. [12] use encryption to protect the parameters of the Isolation Forest models built by the data owners while being shared with a central server. Both aggregated model parameters and thresholds for anomaly detection are calculated by the server using an additive homomorphic encryption scheme and spread to the data owners that will use them locally and decrypted. Liu et al. [19] propose a federated learning approach that combines an attention-based mechanism with LSTM to detect anomalies at the edge, using a cloud aggregator. To reduce the communication overhead generated by the transmission of anomaly detection model's parameters to the cloud aggregator, the authors propose a compression mechanism that also doubles as a security measure.

Finally, decentralized federated learning mechanisms are defined using blockchain networks to record and share trusted information available to all the nodes [16], [20]. Cui et al. [16] propose a system of nodes connected in a P2P network. Each node uses a GAN model to detect anomalies. A differential privacy mechanism protects local models' parameters that are shared with other nodes. The global model aggregation is performed by a node that is chosen using a consensus mechanism implemented on the blockchain. Arazzi et al. [20] propose a system consisting of worker nodes that train local anomaly detection models, aggregators that compute the global detection model and targets representing the monitored devices. A mechanism based on secure multi-party computation is put in place to identify aggregator and worker nodes, while the blockchain is used for implementing a reputation mechanism that estimates the reliability of aggregators and to provide trusted information about malicious nodes.

Table 1 below summarizes the main techniques for privacy preserving anomaly detection highlighting the tradeoffs that need to be considered in implementation.

Table 1. Techniques for privacy preserving anomaly detection.

| IoT data processing | Cryptographic technique | Security Provider | Tradeoffs | |
|---|---|---|---|---|
| | | | (+) | (-) |
| Centralized | <ul><li>Semi-Homomorphic Encryption [3],</li><li>Lightweight Homomorphic Encryption [5]</li><li>Fully Homomorphic Encryption [7][10][11]</li></ul> | <ul><li>Blockchain [7][13][14]</li><li>Random disturbance [3]</li><li>Privacy manager [5]</li><li>Decryptor [10]</li><li>LookUp Table Provider [11]</li></ul> | <ul><li>Ensures data security in transit and in storage.</li><li>Faster and quicker analysis due to cloud resources.</li><li>Easier to design.</li></ul> | <ul><li>Difficult to protect data during processing.</li><li>Decryption is needed.</li><li>Large overhead for computations on encrypted data.</li><li>Large communication overhead due to security protocols put in place for decryption.</li></ul> |
| Federated | <ul><li>Homomorphic Encryption [12][15]</li><li>Secure Multi-party Computation [8][20]</li></ul> | <ul><li>Blockchain [16][20]</li><li>Differential Privacy [17][18]</li><li>Federated Learning [16-19]</li></ul> | <ul><li>Ensures data privacy as it remains with the owner.</li><li>Reduces communication and latency as only processing results are exchanged.</li><li>Computational scalability due to workload distribution.</li></ul> | <ul><li>Interoperability issues due to resources heterogeneity.</li><li>Additional communication overhead introduced by security protocols to protect model parameters exchanged.</li><li>Design complexity.</li></ul> |

Analyzing the state-of-the-art solutions in privacy-preserving anomaly detection, we notice that all existing techniques struggle with reducing communication and computation overhead while maintaining the security of IoT data. The computational overhead is generated by the operations performed on encrypted data. Homomorphic encryption schemes support basic arithmetic operations like addition and multiplication. However, more complex operations such as comparison or division are not supported and require additional cryptographic techniques to be implemented, thus high overheads. Moreover, to avoid this overhead, solutions are decrypting the IoT data from sensors to do analysis and computations, but this exposes the data to privacy risks, as decrypted, raw data is vulnerable to unauthorized access, data breaches, and privacy violations. In this paper, we address the identified knowledge gap in the literature by developing an anomaly detection solution for homomorphic encrypted data generated by IoT devices without relying on partial or total decryption. The proposed solution leverages TFHE scheme implemented in Concrete to work around limited operations on encrypted data and uses a histogram-based technique to identify anomalous sensor readings. The main operations of the Equi-Width Histograms anomaly detection technique including addition, checks on encrypted value membership in buckets, and detection of abnormal bucket counts, were implemented in a vectorized fashion for homomorphic encryption data. Our privacy-preserving anomaly detection technique is robust and efficient for IoT data, taking advantages of the vectorized implementation of the operations, while the computational overheads are kept within reasonable boundaries, like those reported in the literature.

## 3. Fully Homomorphic Encryption

The encrypted homomorphic schemes allow performing additive and multiplicative operations directly on the encrypted data, yielding the same result as if the operations were performed on plain data. Therefore, homomorphic encryption ensures that the sum of two ciphertexts is equal to the cyphertext of the sum of the corresponding plain values and that the product of two cyphertexts is equal to the cyphertext of the product of the corresponding plain values. If $e(x_i)$ is the ciphertext obtained by encrypting the plaintext value $x_i$, homomorphic properties are expressed as follows:

$$e(x_1) + e(x_2) = e(x_1 + x_2) \tag{1}$$

$$e(x_1) \cdot e(x_2) = e(x_1 \cdot x_2) \tag{2}$$

However, despite its promising features for developing privacy preserving solutions in various domains, open challenges limit its application in anomaly detection field. The homomorphic encryption schemes introduce some noise bits in the binary representation of the values to be encrypted to compute the ciphertexts, the number of operations which can be applied on a specific ciphertext is limited, since with each operation the number of noise bits increases. Consequently, the risk of overlapping the noise bits with the actual data bits grows each time a new operation is performed on a particular encrypted value.

Fully Homomorphic Encryption (FHE) allows performing repeated operations on encrypted data without knowing the plain values [36]. Therefore, fully homomorphic encryption allows extending the two fundamental operations and iteratively computing the sum or the product of an arbitrary number of encrypted values, without changing the representation of the actual data being encrypted:

$$\sum_i e(x_i) = e(\sum_i x_i) \tag{3}$$

$$\prod_i e(x_i) = e(\prod_i x_i) \qquad (4)$$

As an unlimited number of operations can be applied on the same cyphertext without altering the actual value being encrypted, FHE is suited for implementing anomaly detection [37]. Furthermore, as not any sequence of arithmetic operations can be reduced to only additions and multiplications, research efforts are concentrated towards extending the range of computationally feasible operations. The goal is to apply a function $f(e(x_1), e(x_2), \ldots, e(x_n))$ on a specific number of encrypted values to obtain an encrypted result without decrypting the arguments. Because implementing any kind of analytical function in FHE is not possible, several approximate implementations which are computationally feasible are defined.

Even though fully homomorphic encryption offers significant advantages for anomaly detection, its support for a limited number of operation types and the constraints regarding the size of the data which can be operated on pose challenges in practical implementations. Therefore, the selection of the appropriate fully homomorphic encryption scheme, as well as the anomaly detection algorithm, is critical for successful deployment in IoT field.

We have opted for the TFHE scheme implemented in Concrete [21][40] as it has several advantages for privacy preserving anomaly detection. It allows translating different types of operations into table lookups, proving considerably higher flexibility in implementing various kinds of data analysis procedures, compared to other FHE schemes that do not support as many operations [39][41][42][43]. Another major feature of Concrete is that it supports function composition [44], allowing to pass the output of a function as input to another function without decrypting the data. This also enables applying repeatedly the same function by forwarding the output of the previous execution as a parameter to a new call of the same function. Furthermore, Concrete offers partial support for floating point operations [45], being able to use floating point values as intermediate results of the computation. However, the Concrete-Compile tool does not support floating point inputs and outputs. Thus, floating point values can be used to compute the result based on the input, but both the input and the output must be integers as all the computation is translated into a table lookup operation which maps integers to an integer result.

Being a fully homomorphic encryption scheme, TFHE implemented in Concrete overcomes the drawback of the traditional encryption schemes in which the data is encrypted just during the transport process between a client and the server which decrypts it, performs the required operations on plain values, encrypts the result and sends it back to the client. In Concrete, the only private key holder is the client, and the server-side operations are entirely performed on encrypted data, using the public evaluation key, which is mathematically related to the client's private key (see Figure 1).

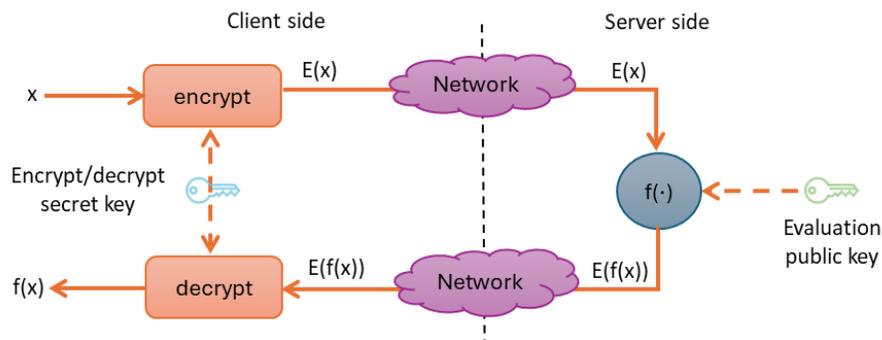

Figure 1. Data encryption and decryption scheme

The data is encrypted using Learning With Errors (LWE) ciphertexts which are bit vectors containing, besides the bits of the actual encrypted data, some random bits, called noise bits. With each operation performed on the ciphertext, the noise part of the representation increases. Thus, to ensure that the noise part of the ciphertext does not overlap with the message part, only a limited number of operations can be performed on each encrypted value. The increase of the noise part can be reduced by using Bootstrap operations which, for a given cyphertext, returns another cyphertext which has the same message part, but less bits for the noise part. After performing a certain number of operations on a ciphertext, a bootstrap operation is needed to reduce the noise.

This representation implies a series of constraints not only on the allowed operations and on the way that programs are written, but also on the size of the plain values. Moreover, this learning-based approach of translating arithmetic operations into table lookup operations is a probabilistic model which is prone to errors. The main drawback of Concrete is that it may not provide the expected answer in all test cases, but its default failure probability is very low:

$$P(f) = \frac{1}{100000} \tag{5}$$

meaning that one execution event of every 100000 may result in an incorrect output.

## 4. Anomaly Detection for Encrypted IoT Data

Due to the limitations of homomorphic encryption in terms of available operations, the anomaly detection on encrypted IoT sensors data streams are challenging. Thus, we have opted for an anomaly detection technique based on Equi-Width Histograms, that uses mainly additive operations and comparisons. This approach circumvents the limitations of homomorphic encryption to enable effective anomaly detection on encrypted data.

Equi-Width Histograms determine the general distribution of the IoT data stream by placing data items in buckets of equal range. In such histograms, buckets have the same size (range), but the number of values that fit in each of them is variable. Each bucket is represented by a tuple:

$$b = (low, high, count), low, high, count \in \mathbb{N} \tag{6}$$

where low and high give the range of values which fit in the bucket and $count$ is the number of elements within the $[low, high)$ range. The algorithm of building an equi-width histogram when the size of a bucket and the minimum and maximum values are known is given below.

---

**Algorithm 1:** Building Equi-Width Histogram

*Input:* $d$ – data sampled during the right operational state of the system, used to build the equi-width histogram

   $s$ – size of a bucket

   $min, max$ – minimum and maximum values

*Output:* $h$ – the equi-width histogram of the data

**Begin**

1:  $h = []$
2:  for i in $min..max - s$ step $s$ do
3:    $b = new\ bucket$
4:    $b.low = i$
5:    $b.high = i + s$

```
6:      b.count = 0
7:      h.append(b)
8:   end for
9:   for each data sample di in d do
10:     for each bucket b in h do
11:        if di >= b.low and di < b.high then
12:           b.count = b.count + 1
13:           break
14:        end if
15:     end for
16:  end for
17:  return h []
End
```

Since anomalies generally have a lower frequency compared to normal data, equi-width histograms can be used at detecting abnormal values by labeling data based on the number of elements within the same bucket. A basic thresholding method can be used to label data which fits in buckets with low count as abnormal. As shown in Figure 1, values which fit into a low-frequency bucket (e.g. 41) are considered to be abnormal, while values which are placed in buckets containing more values (e.g. 62) are labeled as anomalies.

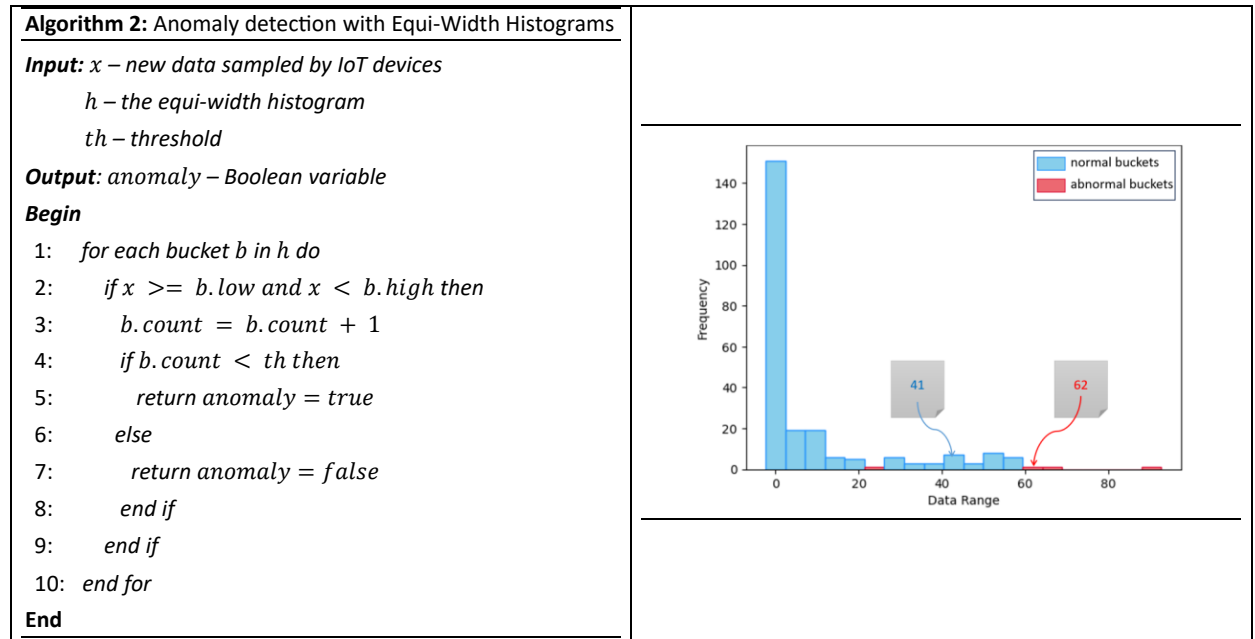

```
Algorithm 2: Anomaly detection with Equi-Width Histograms
Input: x – new data sampled by IoT devices
       h – the equi-width histogram
       th – threshold
Output: anomaly – Boolean variable
Begin
1:   for each bucket b in h do
2:      if x >= b.low and x < b.high then
3:         b.count = b.count + 1
4:         if b.count < th then
5:            return anomaly = true
6:         else
7:            return anomaly = false
8:         end if
9:      end if
10:  end for
End
```

Figure 2. Equi-width histogram basic thresholding for anomaly detection

## 4.1. Equi-width Histograms on Homomorphically Encrypted Data

To address the open challenges for anomaly detection on encrypted data we propose a technique for building equi-width histograms on homomorphically encrypted data, using Concrete library.

First, we define a function which checks if a given value is in a bucket represented by a specific range (see lines 1-2 in algorithm 1). Even though it seems a simple operation it is quite challenging to be implemented directly on homomorphic encrypted data. There are constraints on the number of operations permitted on the same ciphertext value, loops using a counter variable are not permissible, particularly if the number of iterations is large. This limitation arises due to the potential increase in the number of operations performed on the ciphertext value within the loop, which could exceed the allowable limit. As only foreach loops and vectorial operations are allowed without restrictions alternative procedures need to be implemented to ensure compliance with the constraints imposed by the encryption scheme (see algorithm 2 in Figure 3). Therefore, we use the above procedure in a vectorial fashion so that we can check if a given $x$ sensor data fits in a list of buckets, represented as lists of low values, $low[]$, and of high values, $high[]$.

Given an input value $x$ and two lists containing, on the same position, the minimum (low) and, respectively, the maximum (high) value of each bucket, applying the previously defined procedure in a vectorial fashion involves iteratively checking if the input value $x$ fits in the ranged defined by each pair of minimum and maximum values placed at the same index in the range lists. The result of each procedure call is placed on the same position as the limits of the bucket in a new list $r[]$. Therefore, the vectorized version returns a list which is of the same size as the lists for the low and high values and has a single value of 1. All the other elements of the result list are equal to 0. The 1 is placed on the position corresponding to the bucket in which the value fits.

This vectorial approach of determining the bucket in which each value of the input fits can be formally described as a map operation, which takes an IoT data sample and the vectors containing the ranges of buckets as inputs and returns another vector as an output, where each element of the output vector is the result of applying the *is-in-bucket* function to the corresponding element of the input vectors, as follows:

$$is\_in\_bucket : \mathbb{N}^3 \rightarrow \{0,1\}, is\_in\_bucket(x, low, high) = \begin{cases} 1, if\ x \in [low, high) \\ 0, otherwise \end{cases} \quad (7)$$

$$r = vectorial\_is\_in\_bucket(x, \boldsymbol{low}, \boldsymbol{high}) = map\left(is\_in\_bucket, (x, \boldsymbol{low}, \boldsymbol{high})\right) =$$

$$= [is\_in\_bucket(x, low_0, high_0), is\_in\_bucket(x, low_1, high_1), \dots, is\_in\_bucket(x, low_{n-1}, high_{n-1})] \quad (8)$$

where $x$ is a scalar and $\boldsymbol{low}$ and $\boldsymbol{high}$ are lists.

---

**Algorithm 3:** vectorial-is-in-bucket.

**Input:** $x$ - new sampled data by IoT devices
$\quad low[]$ – list of low values of the buckets in histogram
$\quad high[]$ – list of high values of the buckets in histogram

**Output:** $r[]$ – list with a single value of 1 on the position of the bucket in which x fits

**Begin**
1:    r = []
2:    for i in $0..low.length - 1$:
3:        r.append(is-in-bucket(x, low[i], high[i]))
4:    end for
5:    return r []
**End**

---

Figure 3. Procedure determining the bucket where the input value fits in.

Using the vectorial approach to determine the bucket in which the a given value fits, we can compute the histogram of an input data set by adding the result vectors for each value in the input set (see Figure 4). To achieve this, we use the vectorial addition operation which is defined as the element-wise addition of their corresponding components.

---

**Algorithm 4:** Building equi-width histogram from encrypted data.

*Input:* $x[]$ - new sampled data stream from IoT devices
  $low[]$ – list of low values of the buckets in histogram
  $high[]$ – list of high values of the buckets in histogram
*Output:* $h[]$ – Equi-width histogram
**Begin**
 1: $h = [0…0]$
 2: for each xi in x do
 3:  h = h + vectorial-is-in-bucket (xi, low, high)
 4: end for
 5: return h []
**End**

---

Figure 4. Pseudo code for Equi-width histogram building.

For two given vectors $v$ and $u$ of size $n$ the result of the vectorial addition is another vector, $w$, of the same size, defined as follows:

$$\mathbf{w} = \mathbf{u} + \mathbf{v} = [u_0, u_1, \ldots, u_{n-1}] + [v_0, v_1, \ldots, v_{n-1}] = [u_0 + v_0, u_1 + v_1, \ldots, u_{n-1} + v_{n-1}] \quad (9)$$

Generalizing the above relation for the case of m vectors, $\boldsymbol{u}^{(j)}, j = \overline{1, m}$, each having n component, the result of the vectorial addition is:

$$\mathbf{w} = \boldsymbol{u}^{(1)} + \boldsymbol{u}^{(2)} + \cdots + \boldsymbol{u}^{(m)} = [u_0^{(1)}, u_1^{(1)}, \ldots, u_{n-1}^{(1)}] + [u_0^{(2)}, u_1^{(2)}, \ldots, u_{n-1}^{(2)}] + \cdots + [u_0^{(m)}, u_1^{(m)}, \ldots, u_{n-1}^{(m)}]$$

$$= [u_0^{(1)} + u_0^{(2)} + \cdots + u_0^{(m)}, u_1^{(1)} + u_1^{(2)} + \cdots + u_1^{(m)}, \ldots, u_{n-1}^{(1)} + u_{n-1}^{(2)} + \cdots + u_{n-1}^{(m)}] \quad (10)$$

The vectorial addition can be performed only on vectors having the same number of components, since the components of the sum are the element-wise sums of the components of all input vectors.

Therefore, for a given input list of elements, if we perform an element-wise addition of the results given by the vectorial range check function for each element, we determine the equi-width histogram of the input set, as depicted in Figure 5.

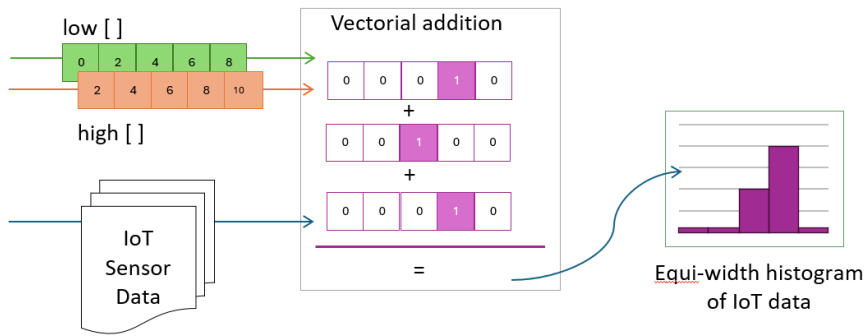

Figure 5. Equi-width histogram building method.

## 4.2. Detecting Anomalies using Encrypted Histograms

In a similar way, we introduce a vectorial approach to the general anomaly detection technique on FHE data which involves labeling data based on a frequency threshold. To fit the constraints on working on encrypted data we are reimplementing lines 4-8 from algorithm 2, in a vectorial fashion. Therefore, we define a procedure that, given an input value and a bucket, checks whether the value fits within the bucket and if the bucket contains fewer elements than a specified threshold value (see Figure 6). By vectorizing the procedure, it can be efficiently applied in a vectorial manner to a set of IoT sensor data.

To achieve this, we define a map operation, which takes an IoT data sample and the vectors containing the ranges of buckets and the number of elements in each bucket and a specific threshold as inputs and returns another vector as an output, where each element of the output vector is the result of applying the *is-in-abnormal-bucket* function to the corresponding element of the input vectors, as follows:

$$is\_in\_abnormal\_bucket : \mathbb{N}^5 \to \{0,1\}$$

$$is\_in\_abnormal\_bucket(x, low, high, count, th) = \begin{cases} 1, if\ x \in [low, high)\ and\ count\ <\ th \\ 0, otherwise \end{cases} \quad (11)$$

$$r = vectorial\_is\_in\_abnormal\_bucket(x, \boldsymbol{low}, \boldsymbol{high}, \boldsymbol{count}, th) =$$

$$= map(is\_in\_abnormal\_bucket, (x, \boldsymbol{low}, \boldsymbol{high}, \boldsymbol{count}, th)) \quad (12)$$

where $x$ and $th$ are scalars and $\boldsymbol{low}$, $\boldsymbol{high}$ and $\boldsymbol{count}$ are lists.

---

**Algorithm 5:** vectorial-is-in-abnormal-bucket.

**Input:** $x$ - new sampled data by IoT devices
  $low[]$ – list of low values of the buckets in histogram
  $high[]$ – list of high values of the buckets in histogram
  $count[]$ – holds the number of elements in each bucket
  $th$ - the threshold for anomaly detection

**Output:** $r[]$ – list with a single value of 1 on the position of the bucket in which $x$ fits

**Begin**
1:   r = []
2:   for i in 0..low.length – 1:
3:       r.append(is-in-abnormal-bucket(x, low[i], high[i], count[i], th))
4:   end for
5:   return r
**End**

Figure 6. Vectorial abnormal value detection in bucket

The vectorial approach to label an input value involves iteratively checking if the value is each of the buckets defined by three lists, containing, on the same position the minimum, the maximum and the number of elements already existing in bucket. When the bucket in which the value must be placed was determined, the number of values that it contains is compared to a given threshold. The result of the vectorial procedure is a list of the same size as the lists containing the range limits for the buckets and it has at most one value of 1. The 1 value is placed on the position corresponding to the bucket that the value can be placed in only if the bucket contains less values than a threshold. Otherwise, all the elements of the resulted list are equal to 0.

To compute the binary label of a specific value, we perform a reduce operation on the result provided by the vectorial-is-in-abnormal-bucket (line 3 in Algorithm 4) procedure by summing all the elements. Reducing a list by summing is a recursive vectorial operation which is defined as follows:

$$reduce(L) = \begin{cases} 0, & L = [] \\ x + reduce(L'), & L = [x] + L' \end{cases} \quad (13)$$

where $L$ is the original list, $x$ is the first element of the list, and $L'$ is the rest of the list. This operation generates a single value that is the sum of all the elements of the list, by adding the first element to the sum of the rest of the list, until the list is empty (see algorithm 5). For an empty list, the reduce by summing operation returns 0.

**Algorithm 6:** abnormal labels in buckets.

**Input:** $x[]$ - new sampled data stream from IoT devices
  $x$ - new sampled data by IoT devices
  $low[]$ – list of low values of the buckets in histogram
  $high[]$ – list of high values of the buckets in histogram
  $count[]$ – holds the number of elements in each bucket
  $th$ - the threshold for anomaly detection

**Output:** $a[]$ – vector of anomalies in buckets

**Begin**
1:  a = []
2:  for each xi in x do
3:    a.append(sum(vectorial-is-in-abnormal-bucket(xi, low, high, count, th))
4:  end for
5:  return a
**End**

Figure 7. Vectorial method to detect outliers using equi-width histograms.

By using this vectorial approach to label data, the actual binary label associated with the input value can be determined by summing all the elements in the list provided by the vectorial procedure. If the list does not contain any value of 1, the sum is 0 and the value is normal. However, if the list contains a value of 1, the sum of all elements is equal to 1 and the value is labeled as abnormal. Therefore, the list of labels is incrementally constructed by appending the label of each value from the input list. The result is a list which contains the binary labels of the input values, placed on the corresponding positions.

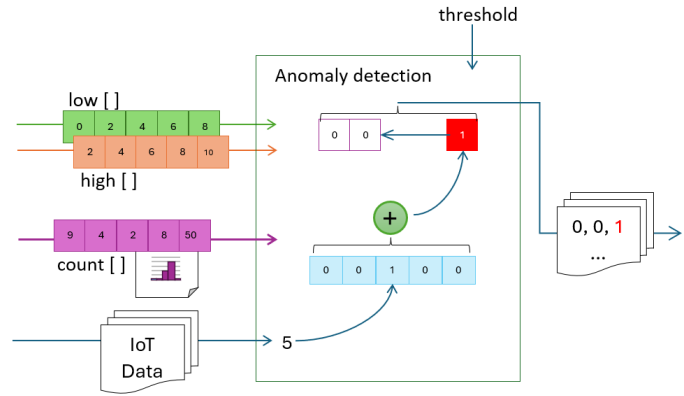

Figure 7. Histogram-based anomaly detection.

## 5. Evaluation Results

To evaluate our prototypes, we use the data provided by IoT devices in a city district, specifically an electricity production dataset [22] containing measurements recorded by smart meters on an hourly basis. Since the values measured by the IoT meters are floating point values between 0 and 9.91 kW and our models can operate only on integer data which has a limited binary representation, we preprocess the electric power data by scaling it to values between 0 and 100 to meet the constraints of our prototypes. To achieve this, we normalize the data to ensure it fits into a limited range, multiply the normalized values by 100 and round the result to the nearest integer, as shown in Figure 8.

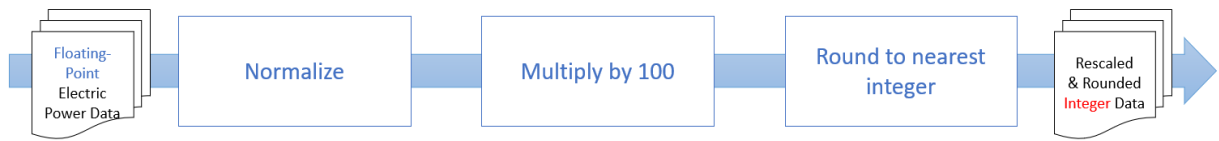

Figure 8. Preprocessing data flow

By applying this technique of data preprocessing which involves data normalization, we aim at converting the data to small sized integers while not disturbing the general distribution of the values.

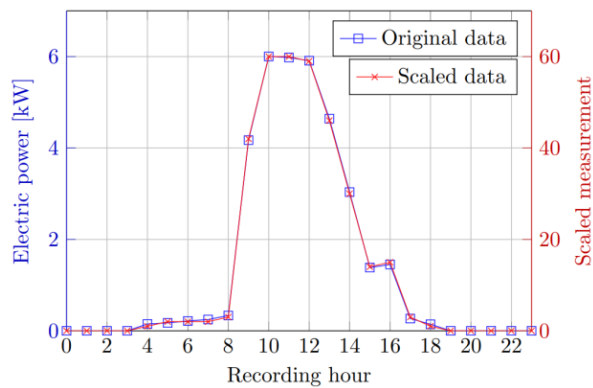

Figure 9. IoT energy meter data before and after scaling

Figure 9 shows the distribution of both the data measured by an IoT energy meter during a specific day before preprocessing and the data which results after normalizing, rescaling the measurements to [0, 100] and rounding the original values to the nearest integers. As the device measures the produced electric power, the values are higher at midday and lower during the morning and the evening. The distribution of the data is preserved by the preprocessing flow that we use, as depicted in Figure 9.

The data has been encrypted using homomorphic encryption and then fed to our anomaly detection method in three configuration cases. We have run several experiments to evaluate the robustness and resilience of the anomaly detection method for encrypted data against various challenges such as noisy sensor data, adversarial attacks, communication failures, and device malfunctions, which often result in sensor reporting a constant value regardless of the conditions to which it is exposed. Therefore, we have artificially introduced anomalies in the preprocessed data by adding noise values to simulate various kinds of abnormal situations which can be encountered in an IoT sensor-based system. To ensure that the data used at building the equi-width histogram reflects the expected behavior of the monitored phenomenon, we consider as the reference data the average of the data recorded by the electric smart meter during five consecutive days.

This histogram-based approach of detecting anomalies based on a frequency threshold performs best on catching anomalies in data containing significant variations and samples which exceed the range of the reference data. However, as the major assumption this technique makes is that anomalies have a considerably lower frequency compared to normal data, this method does not catch constant anomalies when the value reported by the sensor is present within the reference data, used at building the histogram.

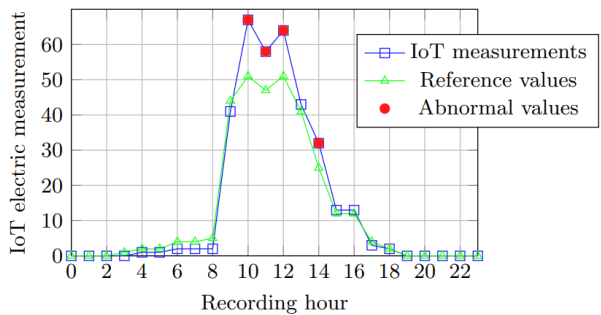

(a) Measurements not matching with the reference histogram

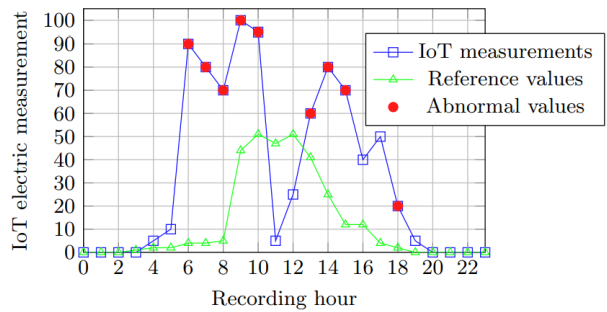

(b) Unusual data pattern with significant variations (spikes)

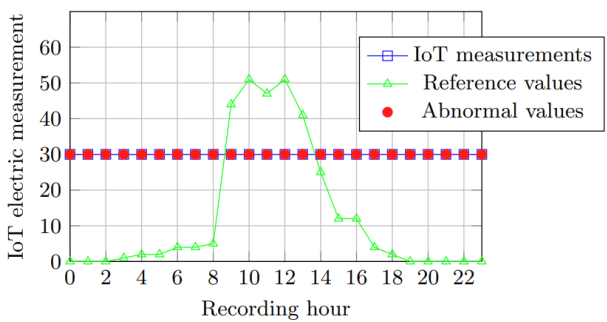

(c) Constant values reported by a faulty IoT smart meter

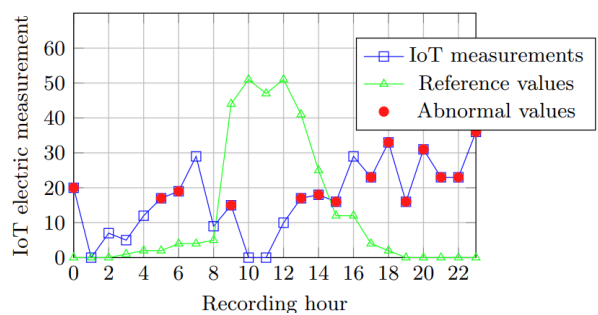

(d) Data produced by a noisy IoT sensor

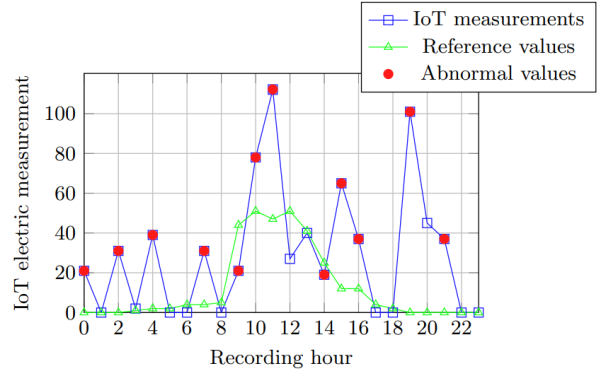

(e) Noisy data exceeding the maximum
value we have considered

Figure 10. Results provided by our privacy preserving anomaly detection method on different kinds of variations which simulate various abnormal situations encountered in an IoT system.

Figure 10 depicts the results of our proposed method on different types of variations, including both real data, measured by the IoT smart meter, and artificially generated data which reflects several types of unexpected behaviors, which may represent abnormal situations. Such abnormalities include malicious attacks, characterized by sudden variations of the measured values, e.g. (b), (e), noisy sensor data, generated by IoT devices which are not properly operating, e.g. (d), (e), communication failures or device malfunctions, e.g. (c).

As presented in Table 2, our results show that the privacy preserving histogram-based anomaly detector provides the same results when operating on homomorphically encrypted data as the non-cryptographic technique on the previously described types of anomalies. Therefore, using encrypted data does not affect the capabilities of the traditional method in terms of anomaly types which can be detected using such an approach.

Table 2. Synthetical description of the results show in Figure 10 on various types of anomalies.

|   | **Anomaly type** | # of anomalies in the dataset | # of normal data samples in the dataset | # of anomalies detected by the non-cryptographic method | # of anomalies detected by the cryptographic method |
|---|---|---|---|---|---|
| (a) | Values not matching with the reference histogram | 3 | 21 | 4 | 4 |
| (b) | Spikes | 12 | 12 | 9 | 9 |
| (c) | Constant | 24 | 0 | 24 | 24 |
| (d) | Noisy sensor data | 22 | 2 | 14 | 14 |
| (e) | Spikes exceeding the considered range | 13 | 11 | 12 | 12 |

## 6. Discussions

In this section we aim to discuss and compare the complexity and execution time of our anomaly detection solution for encrypted data considering three different configuration use cases depicted in Figure 11.

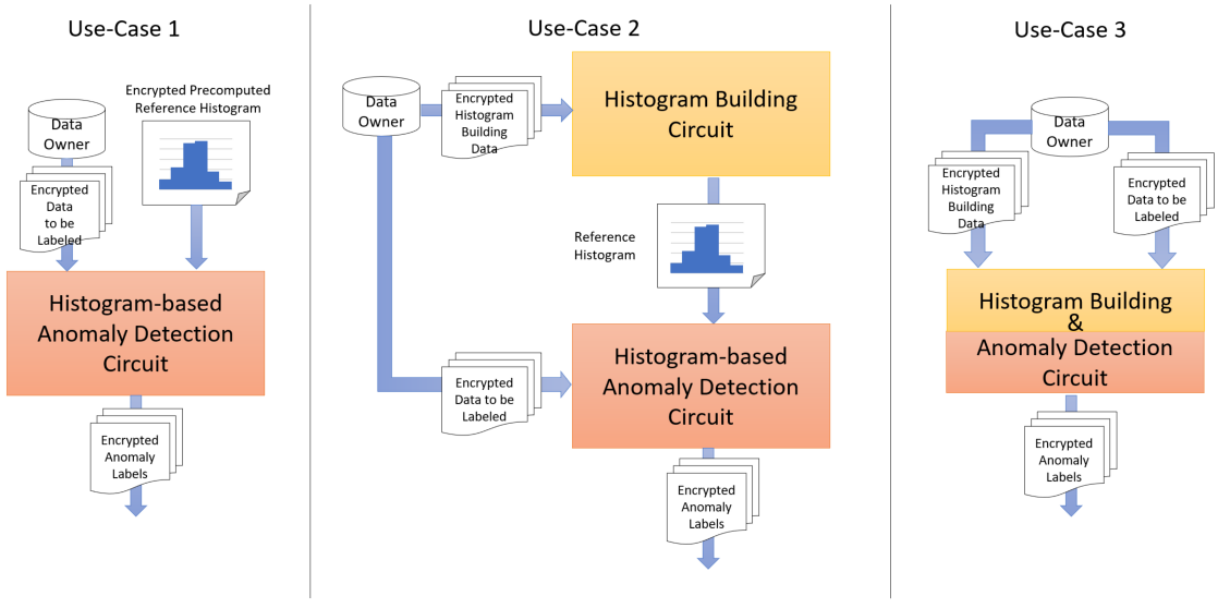

Figure 11. Configuration use case for our anomaly detection solution

The complexity of our configurations is determined in terms of basic operations, such as clear addition, encrypted addition, encrypted negation, and programmable bootstrap, which is used to reduce the amount of noise within the encrypted representation to allow performing multiple operations on the same encrypted operands. The relative complexity of the circuits gives a hint about which circuit is faster (lower complexity will determine a lower execution time) [23]. Table 2 below shows the main differences among the tree configuration use case considered.

Table 3. Configuration of use-cases features.

| Configuration use case | Input (Encrypted) | | | Output (Encrypted) | | # of processing modules |
|---|---|---|---|---|---|---|
| | Data to be labeled | Reference data | Histogram | Anomaly labels | Histogram | |
| UC-1 | ✓ | - | ✓ | ✓ | - | 1 |
| UC-2 | ✓ | ✓ | ✓ | ✓ | ✓ | 2 |
| UC-3 | ✓ | ✓ | - | ✓ | - | 1 |

The first use-case (UC-1 in table 3) assumes the availability of a previously computed histogram of IoT sensor data for the normal behavior of the device. It is used as an input of our anomaly detection circuit, together with the new IoT sensor data to be labeled as normal or with anomalies. In such a configuration case, the expected data variance within the considered time interval is of major importance. Thus, we build the reference histogram from hourly means of energy measurements recorded during several days in the past. UC-1 allows to separately evaluate the performance of the anomaly detection circuit, focusing on catching abnormal values in encrypted data and not on building the data distribution.

We aim to determine how the size of the plain input data samples affects the size of the encrypted inputs, results, and secret keys. As all the values used in computations are preprocessed to fit in [0, 100], they can all be represented using 8 bits. Therefore, when computing the size of the plain inputs, we consider that the size of each value is 1Byte. Thus, since the measurements are recorded on an hourly basis and each bucket is represented by a tuple containing three values, namely $(low, high, count)$, the size of the clear inputs can be expressed, in bytes, as:

$$sizeof(Data) + sizeof(Histogram) + sizeof(threshold) =$$
$$= 24 * No.of\ days + 3 * No.of\ buckets + 1 \qquad (14)$$

Table 4 shows the variation in the size of encrypted inputs, outputs, and secret keys with the size of the plain inputs, expressed as pairs of number of days considered, each containing 24 measurements, and the number of buckets in the histogram based on which the anomaly detection is performed. The actual size of the inputs, in bytes, can be determined using relation 14. The size of the encrypted inputs is correlated to the size of the raw inputs, but the size of the encrypted data is considerably higher. The size of the encrypted output also increases with the size of plain inputs and depends only on the number of measurements to be labeled, as the module outputs an encrypted label associated with each input measurement. However, as can be seen in Table 4, the size of the secret keys is not correlated with the size of input data, and it has the same order of magnitude, regardless of the size of the inputs.

Table 4. UC-1 complexity variation details with respect to the size of inputs, outputs and secret keys.

| Inputs (# days, # buckets) | Encrypted Inputs (B) | Encrypted Outputs (B) | Secret Keys (B) |
|---|---|---|---|
| (1,10) | 12845520 | 295104 | 363808 |
| (1,20) | 18023104 | 295104 | 330336 |
| (1,50) | 34735504 | 295104 | 330400 |
| (2,10) | 19137168 | 590208 | 363848 |
| (2,20) | 24314752 | 590208 | 330392 |
| (2,50) | 41027152 | 590208 | 330448 |

As depicted in Figure 12, even if the size of the encrypted inputs increases almost linear with the size of the clear inputs, the overhead introduced by the encryption is considerably large (x $10^6$).

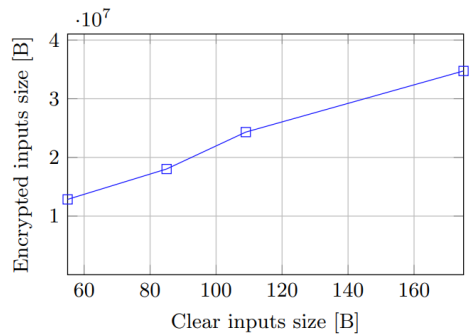

Figure 12. Correlation of the size of encrypted inputs with the size of clear inputs for UC-1.

The model's complexity can also be expressed in terms of operations performed during the anomaly detection process. This complexity for UC-1 is highly dependent on the size of the input samples. As previously described, in our case, the input's length is determined by the number of buckets in the histogram and the number of IoT data samples we want to process. Table 5 summarizes the variation of different types of operations with the size of the inputs. As expected, the number of operations performed increases when the size of the input data increases. However, since we propose an additive algorithm operating on encrypted data, encrypted addition is the most frequently performed operation. The number of programmable bootstrap operations, used to reduce the noise and to allow repeated computations on the same ciphertext, also significantly increases with the size of the inputs as the number of values that must be repeatedly operated grows.

Table 5. UC-1 complexity variation details regarding the operations performed on the encrypted data.

| Inputs (# days, # buckets) | Programmable bootstrap | Key switch | Clear add | Encrypted add | Clear multiply | Encrypted negation | Total operations |
| --- | --- | --- | --- | --- | --- | --- | --- |
| (1,10) | 1859 | 1870 | 1331 | 2891 | 0 | 1859 | 9810 |
| (1,20) | 3549 | 3570 | 2541 | 5541 | 0 | 3549 | 18750 |
| (1,50) | 8619 | 8670 | 6171 | 13491 | 0 | 8619 | 45570 |
| (2,10) | 3707 | 3718 | 2651 | 5771 | 0 | 3707 | 19554 |
| (2,20) | 7077 | 7098 | 5061 | 11061 | 0 | 7077 | 37374 |
| (2,50) | 17187 | 17238 | 12291 | 26931 | 0 | 17187 | 90834 |

A synthetic view of the operations-based complexity variation is depicted in Figure 6. We consider the total number of operations in Table 5 as a measure of the complexity of the model. In this global view, the complexity of our prototype in UC-1 doubles when either the number of buckets or the number of measurements doubles, resulting in an almost linear variation.

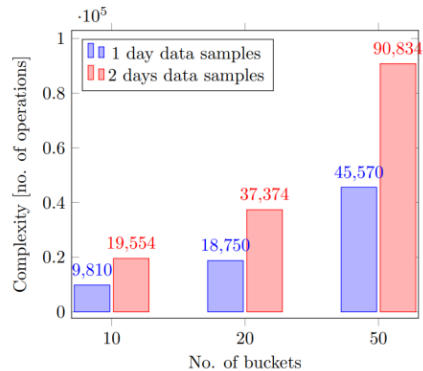

Figure 13. Variation of complexity in configuration UC-1 for different number of buckets and size of input data

We evaluate the performance of our prototype by measuring the compile time which is the time needed to train the model using a set of plain input samples, the time taken by the model to generate the keys and the time taken to generate the encrypted labels associated with the input encrypted measurements. Table 6 shows the time variation for UC-1.

Table 6. UC-1 time performance for different operations

| Inputs (# days, # buckets) | Compile (sec) | Keys Generation (sec) | Execution (sec) |
| --- | --- | --- | --- |
| (1,10) | 97 | 149 | 903 |
| (1,20) | 347 | 116 | 839 |
| (1,50) | 2456 | 135 | 2037 |
| (2,10) | 328 | 68 | 873 |
| (2,20) | 1378 | 66 | 1734 |
| (2,50) | 11419 | 134 | 8290 |

Figure 14 depicts the variation of all time aspects considered for UC-1 configuration, with the size of inputs. The time needed to generate the keys is significantly lower than both compile time and detection (execution) time, which have a similar way of varying the size of input values.

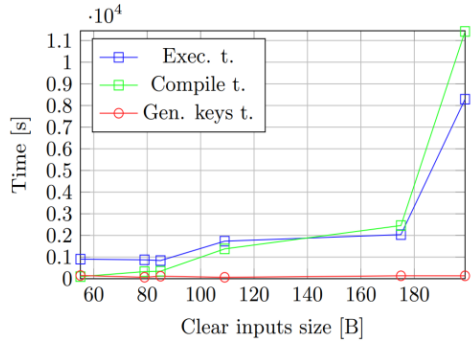

Figure 14. Compile, execution, and key generation times variation for UC-1 with the size of the plain input data.

The second use-case (UC-2 in table 3) aims at individually evaluating the histogram building and the anomaly detection phases of our privacy preservation solution for better capturing the general distribution of the data that may change in time due to seasonality and other external factors. Therefore, the reference histogram must be periodically updated. We build the reference histogram from data reflecting the normal behavior of the system, using a dedicated circuit. This histogram is passed to the anomaly detection circuit together with the data to be labeled. In this way, the two phases of the proposed anomaly detection can be independently analyzed. Passing the reference histogram between two circuits involves decrypting and re-encrypting it using the key of the second circuit or sharing the keys between circuits using a secure third party. The Concrete library allows storing the keys of one on the disk and loading them when building another circuit. This approach, however, exposes the data and induces vulnerabilities which may be easily exploited.

Table 7 presents the complexity with respect to the size of encrypted inputs, outputs and secret keys for both histogram building and anomaly detection modules. The size of encrypted inputs is similar between the two models, while the size of encrypted outputs is significantly higher in the case of the anomaly detection circuit, as it is highly dependent on the size of inputs, consisting of encrypted binary labels corresponding to the input values. Furthermore, the size of the encrypted inputs and outputs of the anomaly detection circuit is like the size of the inputs of the model in UC-1, as we use the same circuit. Considering that during a day 24 measurements are sampled and that the range of each bucket is expressed as a pair of $(low, high)$ values, the size of the clear inputs for the building histogram phase can be expressed as:

$$sizeof(Data_{Ref}) + sizeof(Buckets_{Range}) = 24 * No.of\ days + 2 * No.of\ buckets \quad (15)$$

Table 7. UC-2 complexity variation with respect to the size of inputs, outputs, and secret keys for different phases.

| Phase | Inputs (# days, # buckets) | Encrypted Inputs (B) | Encrypted Outputs (B) | Secret Keys (B) |
|---|---|---|---|---|
| Histogram Building | (1,10) | 12058992 | 180312 | 295240 |
| | (1,20) | 17302032 | 344232 | 295264 |
| | (1,50) | 33031152 | 835992 | 295312 |
| | (2,10) | 18350640 | 180312 | 295264 |
| | (2,20) | 23593680 | 344232 | 295304 |
| Anomaly Detection | (1,10) | 12845520 | 295104 | 363808 |
| | (1,20) | 18023104 | 295104 | 330336 |
| | (1,50) | 34735504 | 295104 | 330400 |
| | (2,10) | 19923600 | 590208 | 429600 |
| | (2,20) | 25035648 | 590208 | 363904 |

The size of the plain inputs of the anomaly detection phase is equal to the size of the inputs in UC-1 (see relation 14). The overhead introduced by encrypting the inputs is around x $10^6$ in the case of both circuits in UC-2, as Figure 15 shows. The variation of the encrypted inputs size with the size of plain inputs is linear for the histogram building prototype and almost linear in the case of anomaly detection prototype.

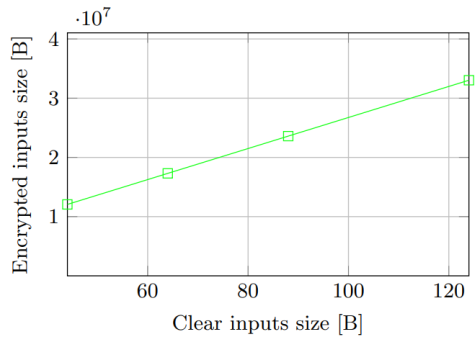

(a) Histogram building phase encrypted inputs size

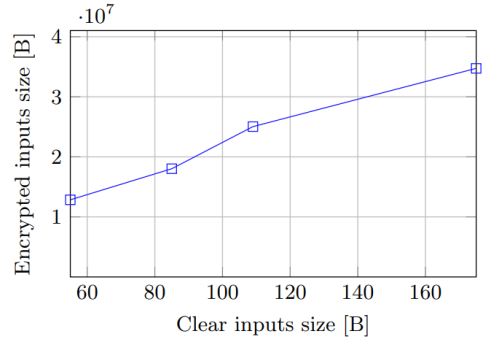

(b) Anomaly detection phase encrypted inputs size

Figure 15. Correlation of encrypted inputs size with the size of clear inputs for both histogram building and anomaly detection phases in UC-2.

A detailed view of the operations-based complexity of the models is presented in table 8, which shows the number of operations of each type performed by both circuits. As in the case of UC-1, the most frequently performed operation is encrypted addition because the fundamental operation for both circuits is vectorial addition. The total number of operations is higher during the anomaly detection phase, resulting in a higher complexity of the circuit, but which is comparable to the complexity of the model in UC-1.

Table 9. UC-2 complexity variation details regarding the operations performed on the encrypted data during different phases.

| Phase | Inputs (# days, # buckets) | Programmable bootstrap | Key switch | Clear add | Encrypted add | Clear mul | Encrypted negation | Total operations |
|---|---|---|---|---|---|---|---|---|
| Histogram Building Phase | (1,10) | 1320 | 1320 | 1056 | 2101 | 0 | 1320 | 7117 |
| | (1,20) | 2520 | 2520 | 2016 | 4011 | 0 | 2520 | 13587 |
| | (1,50) | 6120 | 6120 | 4896 | 9741 | 0 | 6120 | 32997 |
| | (2,10) | 2640 | 2640 | 2112 | 4213 | 0 | 2640 | 14245 |

|  | (2,20) | 5040 | 5040 | 4032 | 8043 | 0 | 5040 | 27195 |
|---|---|---|---|---|---|---|---|---|
|  | (1,10) | 1859 | 1870 | 1331 | 2891 | 0 | 1859 | 9810 |
| Anomaly | (1,20) | 3549 | 3570 | 2541 | 5541 | 0 | 3549 | 18750 |
| Detection Phase | (1,50) | 8619 | 8670 | 6171 | 13491 | 0 | 8619 | 45570 |
|  | (2,10) | 3718 | 3718 | 2651 | 5771 | 0 | 3707 | 19565 |
|  | (2,20) | 7077 | 7098 | 5061 | 11061 | 0 | 7077 | 37374 |

As seen in Figure 16, both the operations complexity of circuits and the difference in the number of total operations performed by our prototype circuits in UC-2 doubles when the size of the data to be labeled or the number of buckets doubles.

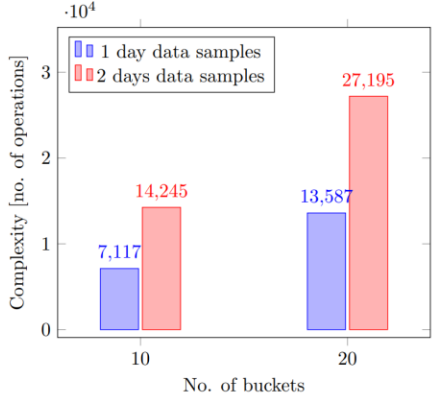
(a) Histogram building phase operations complexity

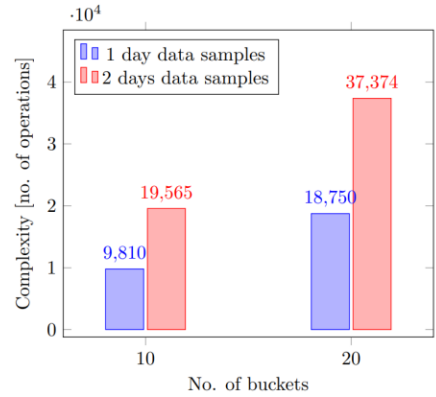
(b) Anomaly detection phase operations complexity

Figure 16. Variation of complexity in configuration UC-2 for different number of buckets and size of input data

The performance of the circuits in UC-2 in terms of compile (training), generating keys and building histogram or detecting anomalies time is synthetically presented in Table 9. The compile time is considerably lower for the building histogram prototype, while the time taken to generate the keys and the execution (building histogram or anomaly detection) time are comparable between the two circuits.

Table 9. Time performance values for UC-2.

| Inputs (# days, # buckets) | Building Histogram | | | Anomaly Detection | | |
|---|---|---|---|---|---|---|
|  | Compile (sec) | Keys Generation (sec) | Execution (sec) | Compile (sec) | Keys Generation (sec) | Execution (sec) |
| (1,10) | 53 | 114 | 813 | 98 | 127 | 824 |
| (1,20) | 215 | 128 | 1734 | 427 | 135 | 1757 |
| (1,50) | 1461 | 129 | 4133 | 3076 | 134 | 4230 |
| (2,10) | 213 | 136 | 1749 | 455 | 170 | 1820 |
| (2,20) | 583 | 61 | 1608 | 1198 | 68 | 1635 |

The execution time generally increases with the size of inputs and, implicitly, with the number of operations performed on input values. However, for both histogram building and anomaly detection circuits, the execution time and key generating time shows a local minimum when the size of inputs is $(\# \, of \, days, \# \, of \, buckets) = (2, 20)$, as depicted in Figure 17, while, in the case of training (compile) time, the local minimum is encountered at a size of inputs of $(\# \, of \, days, \# \, of \, buckets) = (2, 10)$.

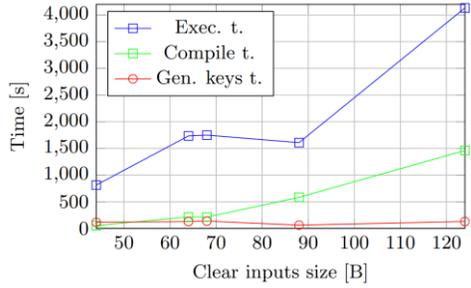 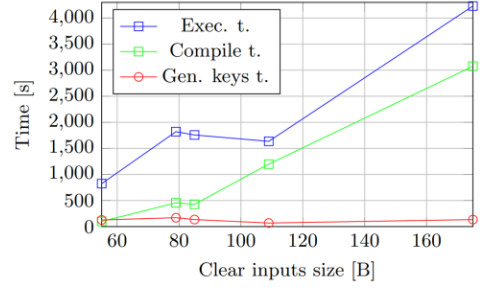

(a) Histogram building phase time variation  (b) Anomaly detection phase time variation

Figure 17. Variation of execution time, compile time, and key generation time for both histogram building and anomaly detection phases of UC-2 with the size of the plain input data

<u>The third use-case (UC-3 in table 2)</u> aims to increase the security of the process, by computing the histogram and detecting anomalies using the same circuit. However, this results in a higher complexity of the circuit and slower execution. Thus, in this scenario, the global complexity of the entire technique is evaluated.

As in the case of UC-1 and UC-2, the complexity of the model regarding the size of encrypted inputs increases with the size of data, while the size of the encrypted outputs is influenced only by the number of buckets in the equi-width histogram. However, as more operations are performed in the model that we define in UC-3, we notice a slight increase in the size of secret keys compared to the size of secret keys generated in UC-1 and UC-2. The total size of the inputs is expressed for UC-3 in a totally similar way to the size of inputs corresponding to UC-1 and UC-2 (see relation 14)

Table 10. UC-3 complexity variation details with respect to the size of inputs, outputs and secret keys.

| Inputs (# days, # buckets) | Encrypted Inputs (B) | Encrypted Outputs (B) | Secret Keys (B) |
| --- | --- | --- | --- |
| (1,10) | 18367032 | 393408 | 367192 |
| (1,20) | 23610072 | 393408 | 367240 |
| (1,50) | 39339192 | 393408 | 334384 |
| (2,10) | 30946232 | 590208 | 428984 |
| (2,20) | 36189272 | 590208 | 429032 |

Since in UC-3 we compute the reference histogram and detect anomalies based on a frequency threshold using a single model, the size of inputs is considerably higher than in the case of circuits defined in UC-1 and UC-2. Therefore, the overhead introduced by encrypting the inputs is even higher than in UC-1 and UC-2, reaching an increase of almost x $2 \times 10^6$ in the size of inputs, as shown in Figure 18.

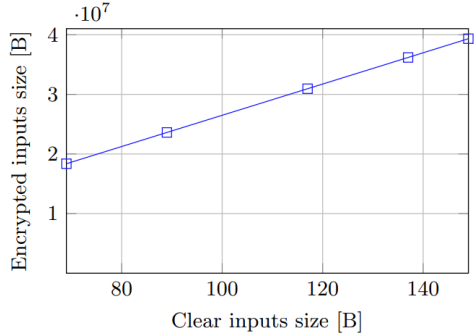

Figure 18. Size of encrypted inputs is correlated to the size of clear inputs for UC-3.

Integrating both histogram building and anomaly detection phases within the same model results in more complex computations performed by the circuit we defined in UC-3. This leads to an inherently higher complexity in terms of performed operations. As Table 11 shows, the number of operations of each type performed by our prototype in UC-3 is approximately equal to the sum of the operations of the same type performed by the circuits which separately build the histogram and detect anomalies in UC-2. Therefore, the complexity of the circuit we use in UC-3 gives an aggregated view of the overall complexity of the privacy preserving anomaly detection technique that we propose.

Table 11. UC-3 complexity variation details regarding the operations performed on the encrypted.

| Inputs (# days, # buckets) | Programmable bootstrap | Key switch | Clear add | Encrypted add | Clear mul | Encrypted negation | Total operations |
|---|---|---|---|---|---|---|---|
| (1,10) | 3179 | 3190 | 2387 | 4992 | 0 | 3179 | 16927 |
| (1,20) | 6069 | 6090 | 4557 | 9552 | 0 | 6069 | 32337 |
| (1,50) | 14739 | 14790 | 11067 | 23232 | 0 | 14739 | 78567 |
| (2,10) | 6347 | 6358 | 4763 | 9984 | 0 | 6347 | 33799 |
| (2,20) | 12117 | 12138 | 9093 | 19104 | 0 | 12117 | 64569 |

As in UC-1 and UC-2, the impact of the number of samples to be labeled and the number of buckets in the reference histogram on the number of operations performed by the model are comparable. Figure 19 shows that doubling the number of data samples and doubling the number of buckets lead to models of comparable operations complexity.

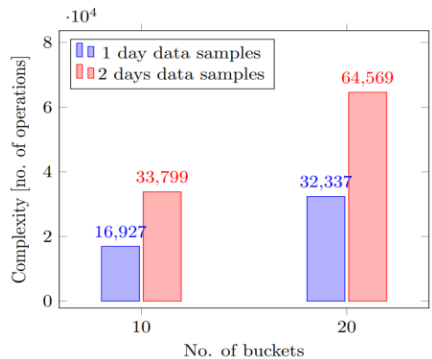

Figure 19. Variation of complexity in configuration UC-3 for different number of buckets and input data samples

The time variation of the model in UC-3 with the size of inputs is summarized in Table 12. Even if the circuit in UC-3 is a composition of the two circuits in UC-2, the compile time, generating keys time and execution time of the circuit is not always equal to the sum of the corresponding time components of the circuits in UC-2.

Table 14. Time variation in UC-3.

| Inputs (# days, # buckets) | Compile (sec) | Keys Generation (sec) | Execution (sec) |
|---|---|---|---|
| (1,10) | 320 | 125 | 1676 |
| (1,20) | 1267 | 128 | 3140 |
| (1,50) | 8265 | 132 | 7645 |
| (2,10) | 917 | 78 | 1657 |
| (2,20) | 3493 | 79 | 3246 |

The execution and compilation time for UC-3 follow the same pattern as the corresponding time components in UC-2. A local minimum can be identified for an input size of $(\#\ of\ days, \#\ of\ buckets) = (2, 10)$ for this prototype. However, taking into consideration the high complexity of the model in terms of both input size and number of operations, the training and execution are slower than in UC-1 and UC-2.

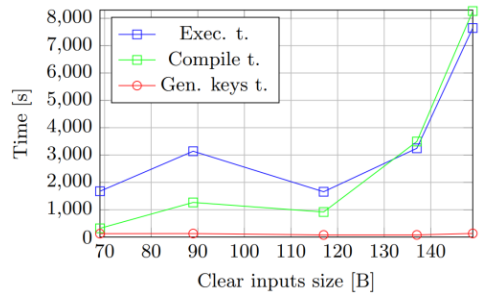

Figure 20. Execution time, compile time, and key generation time for UC-3 with the size of the plain input data

## 7. Conclusion

In this paper we implemented a self-contained anomaly detection solution for homomorphic encrypted data generated by IoT devices. To overcome constraints such as the limited operations that can be performed on encrypted data, we utilized the TFHE scheme implemented in Concrete. This variant of fully homomorphic encryption is specifically designed to be more practical for applications. Our solution employs a redesigned histogram-based algorithm for homomorphic encrypted data streams from IoT devices to identify which sensor readings are statistically different from previous readings, and therefore anomalous. We have implemented the main operations of the Equi-Width Histograms anomaly detection technique adapted for the TFHE homomorphic encryption scheme. The adaptation involves vectorizing additive operations, value placement in buckets, labeling abnormal buckets based on a threshold frequency, labeling abnormal values based on range and corresponding bucket label.

The results obtained are promising, showing that the developed solution can efficiently detect anomalies in fully encrypted data streams without decrypting them. This represents a significant advancement in privacy-preserving in IoT field our solution ensuring the privacy of sensitive data acquired from sensors helping organizations to meet the data protection regulations, such as GDPR in Europe. The execution time overhead introduced by performing operations on encrypted numbers is large, but practical if the processing is done off-line. The labeling of 24 to 48 sensor measurements can take anywhere between a few minutes to a couple of hours, depending on input size and histogram buckets. The exchanged messages between a client and the anomaly detector are only for key distribution and for transferring data to be checked for anomalies. A scenario where the anomaly detector is offloaded closer to the edge is more desirable, since the traffic generated towards the cloud would be considerably less.

The selection of the TFHE scheme and Concrete library for implementing the proposed privacy preserving anomaly solution brings advantages as it supports operations on the encrypted domain, that other similar schemes and libraries do not. Moreover, it generates ready to deploy circuits and provides APIs for easy key management (generate, share and reuse). However, there are still limitations that include constraints on the depth of computation, the precision of arithmetic operations, and the computational complexity of evaluating certain functions.

The results are prone to errors since all the operations performed on homomorphic encrypted data are internally transformed into table lookup operations. The lookup tables are built during a training phase which is performed on unencrypted data and the probability of delivering correct results is highly dependent on the number and the variety of examples used to train the model and the coverage percentage of the possible mappings of input values to results. However, the likelihood of encountering an error due to calculations is very small, much lower than the potential of an IoT device to generate noisy data due to malfunctioning.

Another limitation is the size of data on which operations are performed, this being connected with the overhead associated with performing computations on encrypted data. Specifically, as the IoT sensors' data size increases, the computational complexity of homomorphic operations also increases, leading to longer processing times and higher resource requirements. Our solution considers that the values received from IoT devices are scaled to a predefined range, and the representation must fit in a limited number of bits to avoid overlapping the message part of the representation with the noise bits. This impacts the operations that can be performed on the same value and leads to several constraints in terms of developing anomaly detection solutions. However, vectorial operations are a good and more flexible alternative to traditional iterative operations that are severely restricted. To generate feasible configurations for anomaly detection, in our experiments, small integer numbers (8-bit values) had to be used as inputs.

Finally, the process of designing a sequence of computation on encrypted data is quite inflexible. Any changes to the size of the inputs or the operations that need to be performed will require recompiling the code and retraining the model, these procedures requiring a lot of time (up to several hours). Complex algorithms will result in circuits designed as a composition of multiple functions. This generates considerably larger circuits in terms of the number of basic operations, thus the computations will take more time to execute.

## Acknowledgement

This research received funding from the European Union's Horizon Europe research and innovation program under the Grant Agreement number 101136216. Views and opinions expressed are, however, those of the author(s) only and do not necessarily reflect those of the European Union or the European Climate, Infrastructure, and Environment Executive Agency. Neither the European Union nor the granting authority can be held responsible for them.